\documentclass[12pt]{article}

\usepackage{color}
\usepackage{amsmath}
\usepackage{amsfonts}
\usepackage{amssymb}
\usepackage{caption}
\usepackage{graphicx}
\usepackage{slashed}            
\usepackage[caption=false]{subfig}          
\usepackage{xspace}				
\usepackage{cite}
\usepackage{wrapfig} 
\usepackage{multirow}
\usepackage{float}
\usepackage{placeins} 

\newcommand{\arXiv}[2]{\href{http://arxiv.org/pdf/#1}{{\tt #2/#1}}}
\newcommand{\arXivold}[1]{\href{http://arxiv.org/pdf/#1}{{\tt #1}}}

\newcommand{\Tr}{\mathrm{Tr~}}

\newcommand{\beq}{\begin{equation}}
\newcommand{\eeq}{\end{equation}}

\newcommand{\bea}{\begin{eqnarray}}
\newcommand{\eea}{\end{eqnarray}}

\usepackage[hypertexnames=false]{hyperref}		

\begin{document}
\begin{titlepage}

\vskip.5cm

\begin{center}
{\huge \bf Higgs Decays in Gauge Extensions of the Standard Model} 
\end{center}

\begin{center}
{\bf  {Don Bunk}$^a$, {Jay Hubisz}$^b$, {Bithika Jain}$^b$} \\
\end{center}
\vskip 8pt

\begin{center}

$^a$ {\it  Department of Physics, Hamilton College, Clinton, NY  13323}

$^b$ {\it  Department of Physics, Syracuse University, Syracuse, NY  13244}

\vspace*{0.1cm}

\vspace*{0.1cm}

{\tt  
 
 \href{mailto:jhubisz@physics.syr.edu}{jhubisz@syr.edu},  
 \href{mailto:dbunk@hamilton.edu}{dbunk@hamilton.edu},
 \href{mailto:bjain@syr.edu}{bjain@syr.edu}}
\end{center}

\vglue 0.3truecm

\centerline{\large\bf Abstract}
\begin{quote}


We explore the phenomenology of virtual spin-1 contributions to the $h \rightarrow \gamma \gamma$ and $h \rightarrow Z \gamma$ decay rates in  gauge extensions of the standard model.  We consider generic lorentz and gauge invariant vector self-interactions, which can have non-trivial structure after diagonalizing the quadratic part of the action.  Such features are phenomenologically relevant in models where the electroweak gauge bosons mix with additional spin-1 fields, such as occurs in little higgs models, extra dimensional models, strongly coupled variants of electroweak symmetry breaking, and other gauge extensions of the standard model.  In models where non-renormalizable operators mix field strengths of gauge groups, the one-loop higgs decay amplitudes can be logarithmically divergent, and we provide power counting for the size of the relevant counter-term.  We provide an example calculation in a 4-site moose model that contains degrees of freedom that model the effects of vector and axial vector resonances arising from TeV scale strong dynamics.

\end{quote}

\end{titlepage}

\newpage



\section*{Introduction}
\label{intro}
\setcounter{equation}{0}
\setcounter{footnote}{0}

The discovery of a higgs-like resonance at about $125$ GeV~\cite{CMSHIGGS,ATLASHIGGS} that is so far consistent with expectations from the Standard Model (SM)~\cite{Carmi:2012yp,Azatov:2012bz,Espinosa:2012ir,Giardino:2012ww,Ellis:2012rx}, has altered the landscape of allowed models of electroweak symmetry breaking (EWSB).  The absence of signals in other searches (i.e.~for supersymmetry, and or new resonances) suggests the existence of a gap between the mass of this scalar and other new physics which may be responsible for maintaining the light mass of this scalar field.  A current priority in experimental particle physics is an exhaustive study of this new resonance in terms of a more complete characterization of its production and decays.  

Strong dynamics and/or extra dimensions may still play an important role in protecting the scale of electroweak mass generation from unacceptably large quantum corrections.  The lightness of the higgs could be attributable to it being a pseudo-goldstone boson resulting from the spontaneous breakdown of a global symmetry~\cite{Kaplan:1983sm,Georgi:1984af}, or perhaps conformal invariance of an underlying strongly coupled theory~\cite{Bellazzini:2012vz,Goldberger:2007zk,Fan:2008jk,Chacko:2012vm,Chacko:2012sy}.  It could also be due to geometric warping~\cite{Randall:1999ee,Csaki:2000zn}.  In these cases, the interactions of the light scalar field may be ``higgs-like," although discrepancies relative to the SM predictions generically arise in such theories~\cite{Giudice:2007fh,Barbieri:2007bh,Csaki:2007ns,Low:2009di,Contino:2010mh,Contino:2011np,Farina:2012ea}.  In such cases vector resonances often play an important role in the unitarization of scattering amplitudes of massive SM degrees of freedom~\cite{LlewellynSmith:1973ey,Dicus:1992vj,Cornwall:1973tb,Cornwall:1974km,Lee:1977yc,Lee:1977eg,Chanowitz:1985hj,Csaki:2003dt,Csaki:2003zu,Papucci:2004ip,Foadi:2008xj,Barbieri:2008cc
,Hernandez:2010iu}, and have important phenomenological consequences~\cite{Eboli:2011ye,Bellazzini:2012tv}. 

Extra dimensional solutions to the hierarchy problem predict the existence of a tower of new states beyond those of the SM called Kaluza-Klein (KK) modes.  In such constructions, the gauge bosons of the SM are expected in most models to have corresponding KK-mode partners that appear at energy scales above the inverse size of the extra dimension, along with towers of other spin-1 exotics that are often a key component of such models~\cite{Contino:2003ve,Agashe:2004rs}.

An additional ingredient that may play a vital role in making such theories compatible with other low-energy observables is that of collective symmetry breaking, the mechanism underlying the success of little higgs theories in solving the hierarchy problem~\cite{ArkaniHamed:2001ca,Hill:2000mu,ArkaniHamed:2002qy,Schmaltz:2005ky}.  In these models additional global symmetries, and the particles that complete the SM spectrum into full multiplets of these groups, protect the higgs mass from one- or higher-loop order corrections.  Additional spin-1 states - same spin partners of SM gauge bosons - play a vital role in the cancellation of quadratic divergences in the low-energy effective theory.

In general models of strongly interacting EWSB, including partial UV completions of many little higgs theories, there are also accompanying composite degrees of freedom, beyond those whose masses are protected by spontaneously broken symmetries.  The spectrum of these resonances can be described as a consequence of the pattern of symmetry breaking that occurs below the scale of confinement in a strong sector.  At a minimum, the strong sector must incorporate a custodial $SU(2)$ symmetry in order to protect against unacceptably large contributions to the T-parameter~\cite{Erler:2008zz,Sikivie:1980hm}.  In analogy with QCD, in which the lowest lying vector resonances fit into a representation of the surviving $SU(3)_V$ in the $SU(3)_L \times SU(3)_R \rightarrow SU(3)_V$ chiral symmetry breaking coset, strongly interacting EWSB is expected to at least contain a multiplet of vector resonances fitting into $SU(2)_C$ multiplets resulting from a $SU(2)_L \times SU(2)_R \rightarrow SU(2)_C$ symmetry breaking pattern.  The techniques of effective lagrangians and hidden local symmetry~\cite{Coleman:1969sm,Callan:1969sn,Bando:1984ej,Bando:1987br,Georgi:1989xy,Casalbuoni:1985kq,Casalbuoni:1986vq,Casalbuoni:1995yb,Ecker:1989yg} are particularly convenient methods of parameterizing low energy effective theories that include such vector and/or axial vector resonances.

It is well-known that  higgs production and decay rates can be a bellwether for new physics, especially in the light-higgs window, where numerous channels are available for study.  The majority of higgs events arise from gluon fusion, a one-loop process strongly sensitive to exotic particles with QCD charge which obtain some significant portion of their mass from the higgs mechanism.  In a similar fashion, higgs decays to the di-photon final state are highly sensitive to new particles with non-trivial electro-weak quantum numbers.  In this vein, the hitherto unobserved higgs decay channel $H \rightarrow Z \gamma$ which also occurs only at one-loop order in the SM is another crucially important probe of physics beyond the standard model.  Due to the fact that the rate for the clean final state $l^+ l^- \gamma$ is rather small, the LHC limits are still weak~\cite{Chatrchyan:2013vaa}, and the channel has been a focus of only limited theoretical study~\cite{Gainer:2011aa,Torre:2011bv,Chiang:2012qz,Cai:2013kpa,Azatov:2013ura}.  However, the LHC will soon be exploring the electroweak scale more thoroughly at a center of mass energy scale at or near $13$~TeV.  Of order $100$~fb$^{-1}$ of data are necessary to begin probing the rate expected in the SM, with this luminosity goal achievable in the next couple years of LHC data-taking.

Spin-1 states play a vital role in contributing to the $h\rightarrow Z\gamma (\gamma \gamma)$ channels~\cite{Shifman:1979eb,Gunion:1989we}.  The dominant contribution to both amplitudes in the SM is from virtual $W$ bosons running in loops, with virtual top quarks giving the next largest piece of the amplitudes.  In extensions of the SM, the higgs-$WW$ coupling is often modified, generating corrections to these amplitudes.  Exotic spin-1 states also appear in numerous constructions (such as those described above) and should give contributions at one-loop as well.  In this work, we study generic virtual spin-1 contributions to higgs decays, using the most general set of vector self-interaction terms consistent with $U(1)_\text{EM}$ gauge invariance.  We have calculated skeleton amplitudes that we have made available as Mathematica readable files for use by those wishing to calculate such amplitudes in their model of choice~\cite{mathematicafiles}.  We exhibit the utility of these amplitudes in the context of an explicit moose construction with resonances that model vectors and axial vectors in strongly coupled extensions of the SM that preserve a custodial $SU(2)$ symmetry.  The model, which is a modification of the construction detailed in~\cite{SekharChivukula:2008gz} with the addition of a higgs-like resonance, exhibits the full range of possibilities for the couplings associated  with vector self-interactions.  Additionally, the model incorporates a dimension-6 operator with a coefficient whose value affects the $S$-parameter (which is typically large in models where strong dynamics plays a role in electroweak symmetry breaking~\cite{Holdom:1990tc,Golden:1990ig,Peskin:1991sw,Barbieri:2004qk,Cacciapaglia:2006pk
,Orgogozo:2011kq}).

The organization of this paper is as follows.  In Section~\ref{sec:generalities}, we describe a framework for constructing gauge invariant low-energy effective theories that allow for modified higgs couplings to both SM and exotic spin-1 states, and also allow for a complete range of vector cubic and quartic self-interactions consistent with $U(1)_\text{EM}$.  In Section~\ref{sec:diagrams}, we describe the relevant Feynman rules in a generic framework, and outline our parameterization for the one-loop amplitudes.  In Section~\ref{sec:VandAVres}, we construct an explicit model in which we derive the Feynman rules relevant for a calculation of the $h\rightarrow \gamma \gamma (Z\gamma)$ amplitudes.  In Section~\ref{sec:results}, we explore the decay rates over the parameter space of the model, paying particular attention to correlations between the tree-level contribution to the S-parameter, the $h\rightarrow \gamma\gamma$ rate, and the $h\rightarrow Z\gamma$ rate as these are of especial interest in these types of effective theories~\cite{Gillioz:2012se,Grojean:2013kd}.  We conclude in Section~\ref{sec:conclusions}.   Mathematica files containing skeleton amplitudes (and couplings for our explicit calculation) that can be used in generic gauge extensions of the SM can be downloaded online~\cite{mathematicafiles}.


\section{General Vector Interactions}
\label{sec:generalities}
Diagonalization of the quadratic part of actions that arise in gauge extensions of the standard model often result in mixing of the SM vector fields with exotic ones.  This mixing results in shifted gauge boson self interactions such that the $W$, $Z$, and higgs boson couplings differ from those of the SM.  In addition, the light fields will also generically have direct couplings to heavy exotica.  The higgs boson couplings to the gauge fields will also depend on how the observed scalar higgs is embedded into the complete mechanism of gauge symmetry breaking, including both electroweak breaking and the breaking of the extended gauge sector.  In this section, we describe the classes of actions we consider, and we then characterize the most general self-interactions of the vector fields with each other and with the higgs, under the constraint that all interactions be gauge invariant.
\subsection{The Quadratic Action}

We consider a generic gauge group $G$ with a kinetic term constructed from the usual gauge-invariant field strengths:
\begin{equation}
\label{eq:wfmixing}
\mathcal{L}_{\mathrm{kin}}=-\frac{1}{4}  \Tr V_{\mu\nu} V^{\mu\nu}.
\end{equation}
The trace in this equation is over all generators of the UV gauge group, and
at a minimum, this complete gauge group must contain the electroweak group $SU(2)_L \times U(1)_Y$, either trivially as a product structure, or embedded into a higher rank group.

To describe the breaking of this gauge group down to  $U(1)_\text{EM}$, we construct a low energy effective theory in which complete gauge invariance is realized non-linearly in an effective field theory.  The gauge symmetry breaking of the extended gauge sector can be parameterized by a set of $\Sigma$-fields whose vacuum expectation values determine the spectrum.  The mass terms for the spin-1 fields arise from a sum over the kinetic terms for these $\Sigma$-fields:
\begin{equation}
\label{eq:vectormassterms}
\mathcal{L}_\text{mass}= \sum_{l} \frac{f^2_l}{4} \Tr \vert D_l^{\mu} \Sigma_{l}\vert ^2.
\end{equation}
Mass mixing between the gauge eigenstates arises from these kinetic terms when the sigma fields are expressed in terms of their vacuum expectation values $\Sigma_l \rightarrow \Sigma^0_l$.  These vev's are taken such that the desired breaking pattern $G \rightarrow U(1)_\text{EM}$ is obtained.

In the spirit of low energy effective theory, we should consider terms involving additional insertions of the $\Sigma$ fields that may contribute to the low energy effective action.  Such operators are non-renormalizable, and should be thought of as the product of having integrated out some UV dynamics, which may be either strongly or weakly coupled.  The most phenomenologically interesting class of operators from the standpoint of electroweak precision or contributions to higgs decay phenomenology is the addition of wave function mixing operators:
\begin{equation}
\label{eq:wfmixing}
\mathcal{L}_{\mathrm{WF}}=\epsilon_{ij} \Tr V^i_{\mu\nu} \Sigma_{ij} V^{j \mu\nu} \Sigma^\dagger_{ij}.
\end{equation}
Such operators are closely analogous with the operator that corresponds to integrating out UV dynamics which contributes to the oblique S-parameter:
\begin{equation}
{\mathcal O}_S = \frac{1}{\Lambda^2} H \tau^a H^T W_{\mu\nu}^a B^{\mu\nu}.
\end{equation}
In fact such operators, with properly chosen coefficients, can contribute to a reduction in the severity of electroweak precision constraints in models of vector and axial-vector resonances such as those that appear in extra dimensional models of electroweak symmetry breaking, in strongly coupled UV completions of little higgs models, and generically in various $L-R$ symmetric variants of gauge extensions of the SM.  We further discuss the correlations between electroweak precision observables and the higgs decay rates in Section~\ref{sec:results}.

\subsection{Vector boson self-interactions}
The most general set of 3-point interactions involving two charged vectors with a neutral one (here displaying only the $\gamma$ or the $Z$), that are consistent with conservation of electric charge, are given by:
\begin{align}
{\mathcal L}_{3} = -i \sum_{X,Y} & g_\gamma^{X,Y} X^+_\mu Y^-_\nu A^{\mu\nu}  + g_Z^{X,Y} X^+_\mu Y^-_\mu Z^{\mu\nu} + G_Z^{X,Y} \left( X^+_{\mu\nu} Y^{-\mu}- X^-_{\mu\nu} Y^{+\mu}\right) Z^\nu \nonumber \\
&+ e \left(X^{+\mu\nu} X^-_{\nu} - X^{-\mu\nu} X^+_{\nu} \right) A_\mu
\end{align}
where $X$ and $Y$ are vector fields.  These may be either SM $W^\pm$ bosons, or exotic vector resonances.   The first three of these terms clearly transform trivially under electromagnetic gauge transformations, while the 4th manifests gauge invariance only after considering transformations of the quartic interactions.   ${\mathcal L}_3$ thus contains interactions of the SM gauge fields with each other with possibly modified coupling values and interactions of exotic charged states with the photon and $Z$.  While the last coupling is fixed by gauge invariance, the others are free parameters up to inter-relations arising from the need to preserve full gauge invariance of the complete UV gauge structure giving rise to these interactions.

Similarly, the 4 point interactions take the form\footnote{Up to interactions with more than two derivatives such as $\frac{1}{\Lambda^4} F^4$ non-renormalizable operators, where $F$ is the field strength corresponding to the spin-1 fields in the effective theory.} 
\begin{align}
{\mathcal L}_4 = -&\sum_{X,Y}  A_\mu Z_\nu X^+_\rho Y^-_\sigma \left( 2 a_{\gamma Z}^{XY} g^{\mu\nu} g^{\rho\sigma}- b_{\gamma Z}^{XY} g^{\mu\rho} g^{\nu\sigma}  - c_{\gamma Z}^{XY} g^{\mu\sigma} g^{\rho\nu} \right) \nonumber \\
 &-  \sum_X e^2 A_\mu A_\nu X^+_\rho X^-_\sigma \left( 2 g^{\mu\nu} g^{\rho\sigma}-  g^{\mu\rho} g^{\nu\sigma}  -  g^{\mu\sigma} g^{\rho\nu} \right).
\end{align}
For the mixed $\gamma Z$ coupling, the coefficients are determined by the requirement of overall gauge invariance of the full theory.  Electromagnetic gauge invariance forces the $\gamma \gamma$ quartic couplings to be equal to the square of the electromagnetic coupling constant.

\subsection{Higgs interactions}
For the purposes of this paper, we maintain a semi-model independent attitude regarding the origin of the observed higgs-like scalar field.  We take an effective field theory approach, assuming the higgs is a CP even singlet under electromagnetism, and we allow its couplings to the various vector fields to be free parameters.  Inspired by the higgs low-energy effective theorems, in which the SM higgs interactions are derived (in the approximation that $p_H \rightarrow 0$) by substituting occurrances of the weak scale vacuum expectation value with $v \rightarrow v (1 + h/v )$, we scale all non-linear sigma model vev's by $f_i \rightarrow f_i (1 + a_i h/f_i)$, where the $a$'s are free parameters of the low energy effective theory.  Applying this formalism to the $\Sigma_l$ kinetic terms in Eq.~(\ref{eq:vectormassterms}), we have
\begin{equation}
\label{eq:hgWF}
\mathcal{L}_{\mathrm{h-V}}= \sum_{l} \left( 2 a_l\frac{h }{f_l}\right) \frac{f^2_l}{4} \Tr \vert D_l^{\mu} \Sigma_{l}\vert ^2.
\end{equation}
Specific models will generate different values for these coefficients, although there are constraints from requiring  perturbative unitarity of the effective theory~\cite{Bellazzini:2012tv}.  Particularly, they need not be ${\mathcal O}(1)$, and indeed can be much smaller in some models.

\section{Diagrams}
\label{sec:diagrams}
\begin{figure}[h]
\center{\includegraphics[height=1.1in]{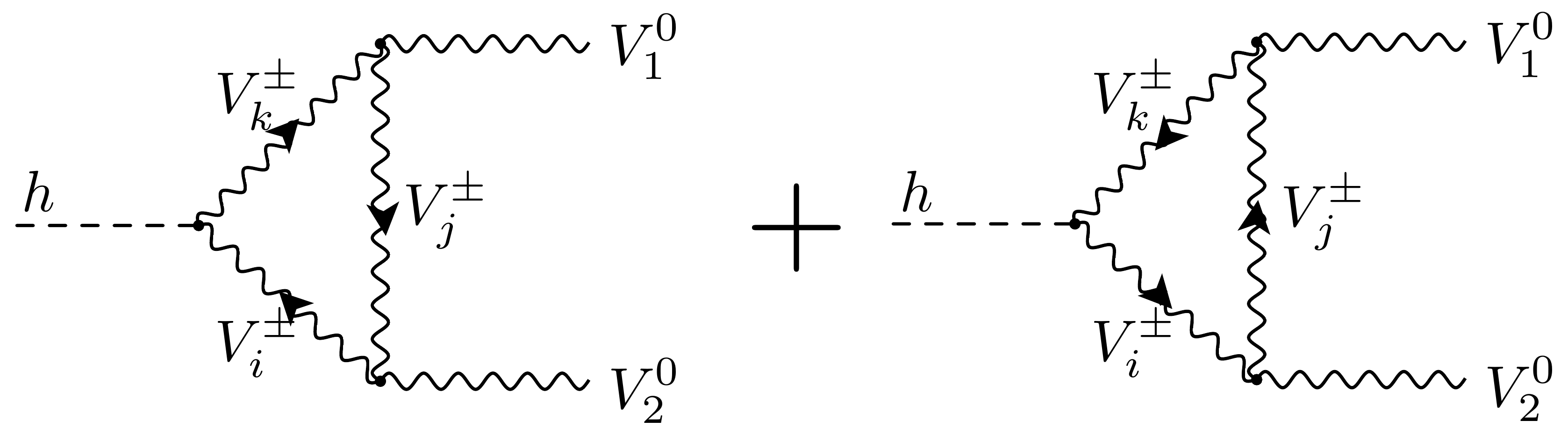}\includegraphics[height=1.1in]{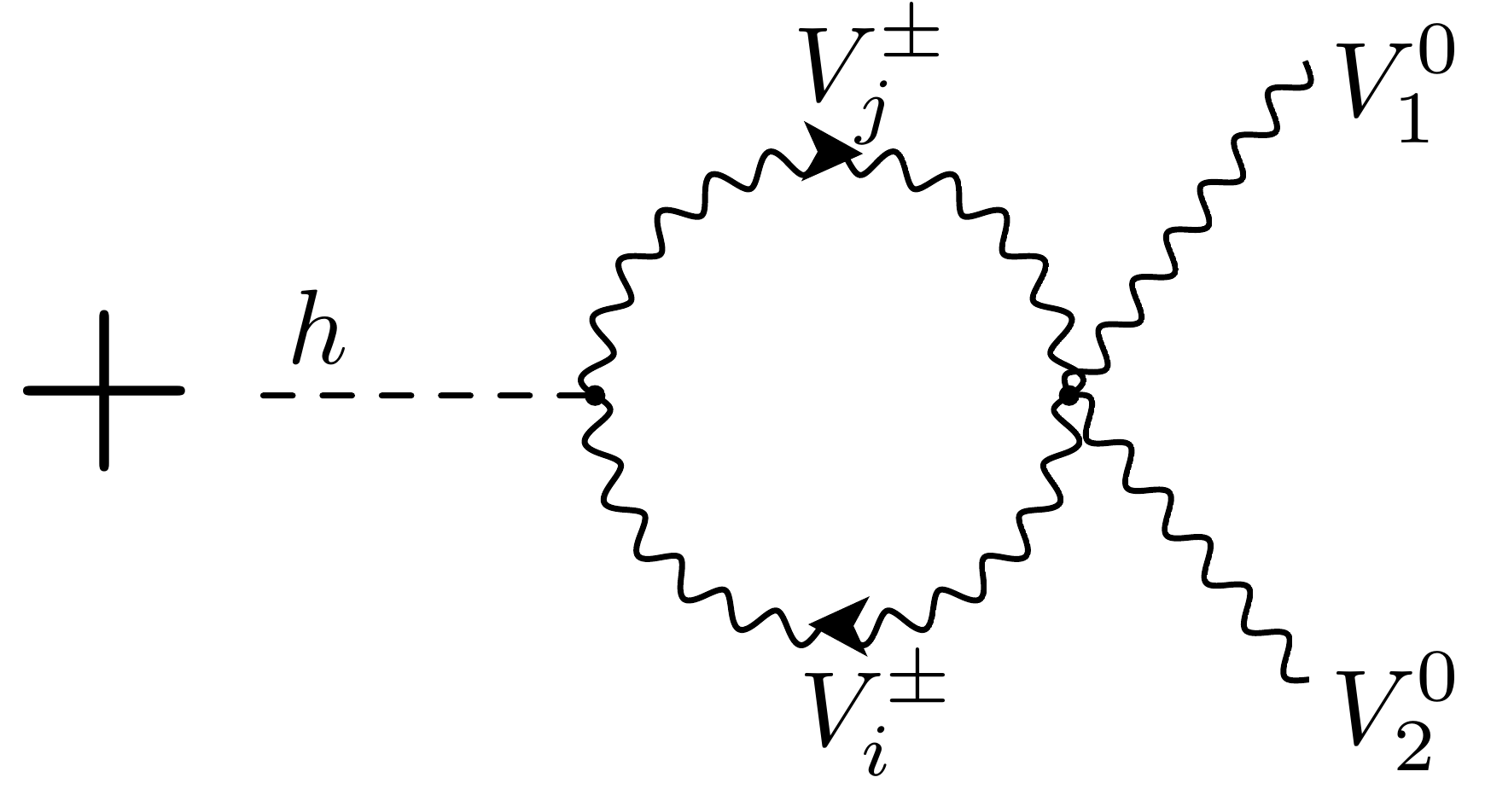}}
\caption{One-loop diagrams contributing to the scalar decay rate to neutral vector bosons (i.e. the photon or $Z$) in gauge extensions of the SM.  There is an implied summation over all charged spin-1 fields in the model.  The arrows on the charged vector field propagators indicate direction of charge flow.  We refer to the sub-amplitudes corresponding to these diagram types as $\mathcal{A}$, $\mathcal{A}^\times$, and $\mathcal{A}^\propto$, respectively.}
\label{fig:diagrams}
\end{figure}

We have computed the most general possible diagrammatic structure for loop processes involving the contributions of virtual spin-1 fields to effective operators coupling the scalar higgs to field strengths for vector bosons.  The loops of consequence in amplitudes for $h \rightarrow \gamma \gamma (Z \gamma)$ involve charged vector bosons running in loops through both ``triangle" and ``fishing" diagrams, shown in Figure~\ref{fig:diagrams}.

\begin{figure}[h]
\center{\includegraphics[width=4in]{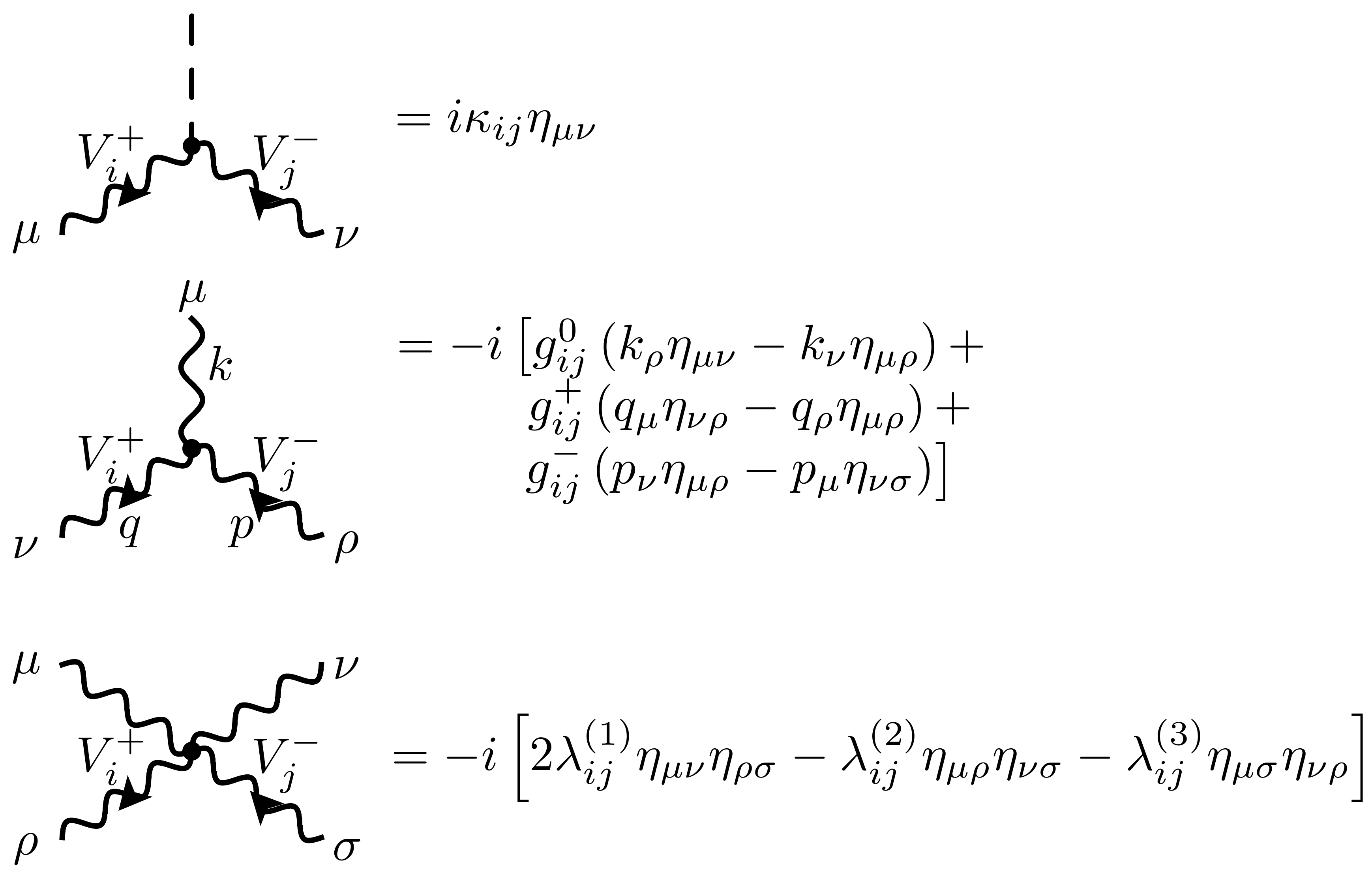}}
\caption{Feynman rules for vertices with general interaction structure.  Rules for vertices with 2 charged particles are shown, as these are what are relevant for the calculation.  All momenta are assumed to be entering the vertices, and arrows indicate charge flow.}
\label{fig:frules}
\end{figure}

As discussed in Section~\ref{sec:generalities}, the vertex structure for the interactions is non-standard in generic models, and we characterize the Feynman rules relevant for the computation in Figure~\ref{fig:frules}.

With the assistance of the FeynCalc package for Mathematica~\cite{Kublbeck:1992mt}, we have calculated the diagrams corresponding to the range of possible vertex structures shown in Figure~\ref{fig:frules} by turning on one form of coupling at a time.  For the triangle diagrams, this corresponds to computing $3\times3$ matrices of amplitudes, $[\mathcal{A}(M_i^2,M_j^2,M_k^2)]_{\alpha\beta}$ and $[\mathcal{A}^\times (M_i^2,M_j^2,M_k^2)]_{\alpha\beta}$, for each of the possible charge flow directions, taking each vertex to have one of $g^0, g^+,$ or $g^-$ set to one, with all others turned off.  For the fishing diagrams, we compute a vector of diagrams, $[\mathcal{A}^\propto (M_i^2,M_j^2)]_\alpha$, with each of the $\lambda^{(1,2,3)}$ couplings set to one, the others to zero.  These individual amplitudes are divergent, and we report the finite and divergent parts of these diagrams (computed in unitary gauge) in an online repository of Mathematica files~\cite{mathematicafiles}.  The full amplitude in a specific model is then obtained by contracting these arrays of sub-diagrams with arrays of couplings that are specific to a given model.
The summation over virtual spin-1 fields and their associated couplings to the higgs and external neutral gauge fields is given by:
\begin{align}
\label{eq:ampsum}
&\mathcal{M}^{\mu\nu}_{ V^1_0 V^2_0} = \sum_{i j k \alpha \beta} [\kappa_h]_{k i} [\mathcal{A}^{\mu\nu} (M_i^2,M_j^2,M_k^2) ]^{\alpha \beta} [g_{V^1_0}]^\alpha_{ji} [g_{V^2_0}]^\beta_{kj} \nonumber \\
& \mathcal{M}^{\times\mu\nu}_{ V^1_0 V^2_0}= \sum_{i j k \alpha \beta} [\kappa_h]_{i k} [\mathcal{A}^{\times\mu\nu} (M_i^2,M_j^2,M_k^2)]^{\alpha \beta} [g_{V^1_0}]^\alpha_{ij} [g_{V^2_0}]^\beta_{jk} \nonumber \\
& \mathcal{M}^{\propto\mu\nu}_{ V^1_0 V^2_0} = \sum_{i j \alpha } [\kappa_h]_{j i} [\mathcal{A}^{\propto\mu\nu}(M_i^2,M_j^2)]^\alpha  [\lambda_{V^1_0 V^2_0}]^\alpha_{ij},
\end{align}
where $V_0^1$ and $V_0^2$ are the external neutral gauge fields, either $\gamma$ or $Z$. 
These contributions must then be summed together to obtain the full amplitude due to spin-1 contributions.

In the next sections we explore contributions to the higgs partial widths in a specific extension of the SM that exhibits the full generality of the couplings and diagrams that have been discussed thus far.


\section{A Specific Model:  \\Vector and Axial-Vector Resonances}
\label{sec:VandAVres}
If the 126 GeV resonance is produced as a composite of TeV scale strong dynamics, it is likely that there are a host of other composite states with masses not far above the electroweak scale.  These states should occupy representations of the symmetries of the UV theory.  The approximate $SU(2)_L \times SU(2)_R$ global symmetry of the low energy theory, which enforces the absence of tree-level corrections to the oblique $T$-parameter, dictates that the symmetries of the UV should reflect at least this global symmetry, with a spontaneous breaking pattern  $SU(2)_L \times SU(2)_R \rightarrow SU(2)_V$, mimicking the custodial symmetry breaking pattern of the SM.  In analogy with QCD, there may be vector and axial vector states, transforming non-linearly as the broken and unbroken generators for these symmetries.  The $SU(2)$ structure implies that these states should fall into triplets with a charged and neutral vector in each:  $\rho^{\pm,0}_V$ and $\rho^{\pm, 0}_A$.  If these states are light in comparison with the scale associated with non-perturbativity of the effective theory, then they can enter in loop processes and give calculable contributions to the effective interactions of the scalar resonance.

Axial vector resonances are especially interesting from the perspective of electroweak precision due to the fact that their contribution to the $S$-parameter can partially cancel contributions from the vector resonances~\cite{Orgogozo:2011kq}.  In this section, we study the couplings of charged vector and axial vectors relevant for the higgs decay rates to $\gamma\gamma$ and $\gamma Z$ in the context of a model which thoroughly explores the range of possibilities for exotic gauge boson self-interactions.


\subsection{Effective Lagrangian for Vectors and Axial Vectors}\label{Ela}
\begin{figure}[h]
\center{\includegraphics[width=4in]{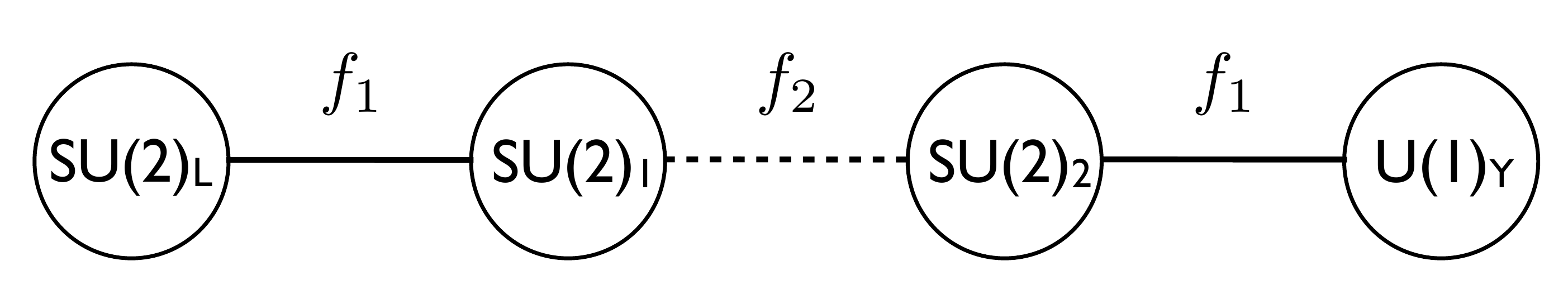}}
\caption{The moose diagram that we study that incorporates vector and axial vector resonances}
\label{fig:moosefig}
\end{figure}

A completely generic implementation of axial vector resonances is difficult from the perspective of the low energy effective theory in which only transformations under the unbroken global $SU(2)_C$ are invariants of the phenomenological Lagrangian~\cite{Coleman:1969sm,Callan:1969sn}.  Axial vectors can be implemented in different ways while remaining consistent with the unbroken $SU(2)_C$~\cite{Bando:1987ym,Casalbuoni:1988xm}.  To have a concrete model to study, which should have some features of actual strongly coupled theories while also allowing concrete results from computation, we study the theory described by the moose diagram shown in Fig.~\ref{fig:moosefig}.  We note that this is precisely the moose studied in~\cite{SekharChivukula:2008gz}, although we are considering the effects of adding a singlet $h$ to these models which is coupled in a delocalized way to gauge fields.

The model incorporates a $SU(2)_1 \times SU(2)_2$ gauge extension of the SM, with degrees of freedom referred to as vector and axial vector triplets of resonances.   We impose a $L - R$ symmetry to preserve custodial $SU(2)$.   This $L-R$ symmetry forces the link vevs between $SU(2)_L - SU(2)_1$ and $SU(2)_2 - U(1)_Y$ to be equal - both are given by $f_1$.  Additionally imposing the parity symmetry requires that the couplings associated with $SU(2)_1$ and $SU(2)_2$ be equal - in our case $g_1 = g_2 \equiv g_\rho$. This $P_{LR}$ is broken explicitly by the SM hypercharge interactions, as $U(1)_Y$ corresponds to gauging only the $t_3$ generator of $SU(2)_R$.  This is the usual case in the SM, where it is the hypercharge interactions (as well as the fermion Yukawa couplings) that violate custodial symmetry.  In writing the action for this theory, we take the usual canonically normalized gauge kinetic terms for the $4$ gauge groups:
\begin{equation}
 \mathcal{L}_{\mathrm{gauge-kin}} = -\frac{1}{4} \left[W_{\mu \nu}^{a~2} + X_{(1) \mu \nu}^{a~2} + X_{(2)\mu \nu}^{a~2} +B _{\mu \nu}^2\right].
\end{equation}

Strongly coupled models of electroweak symmetry breaking are commonly afflicted by severe electroweak precision constraints, even with a custodial symmetry imposed.  Finding models in which the oblique $S$-parameter is small is the biggest challenge~\cite{Holdom:1990tc,Golden:1990ig,Peskin:1991sw}.  Generically, tree level contributions to the $S$-parameter arise from mixing of the vector and axial vector states with the SM gauge fields.  In~\cite{SekharChivukula:2008gz}, it was shown that it is possible to reduce the $S$ parameter with a higher dimensional operator that kinetically mixes $SU(2)_1$ and $SU(2)_2$, analogous with a similar technique in holographic technicolor models~\cite{Hirn:2007we}.  As discussed above, this kinetic mixing gives rise to non-trivial structure for the interaction vertices for the gauge fields once the quadratic Hamiltonian is diagonalized.  The gauge invariant kinetic mixing term we consider is given by
\begin{equation}
\mathcal{L}_{\mathrm{WF}} = -\frac{1}{2} \epsilon~ \text{Tr} \left[ X_{(1) \mu \nu} \Sigma_{12} X_{(2)}^{\mu \nu} \Sigma_{12}^{\dagger}\right],
\label{eq:WFterm}
\end{equation}
where $\Sigma_{12}$ is the nonlinear sigma model link field corresponding to the central line connecting the $SU(2)_1$ and $SU(2)_2$ gauge groups in the moose.  The spin-1 cubic and quartic interactions arise from both the standard and wave-function mixing kinetic terms.  Note that the parameter $\epsilon$ must be constrained $-1 < \epsilon < 1$ to avoid ghosts in the field theory, and that there are limits which strongly imply that $S$ must remain positive~\cite{Agashe:2007mc}.

The gauge kinetic terms for the $\Sigma$-fields determine the structure of the mass matrix for the gauge fields.  We consider the following Lagrangian for these gauge kinetic terms:
\begin{equation}
{\mathcal L}_{\Sigma-\text{kin}}  = \frac{f_1^2}{8} \text{Tr} \left[ \left| D_\mu \Sigma_{L1} \right|^2 \right] +  \frac{f_2^2}{8} \text{Tr} \left[ \left| D_\mu \Sigma_{12} \right|^2 \right] + \frac{f_1^2}{8} \text{Tr} \left[ \left| D_\mu \Sigma_{2Y} \right|^2 \right],
\label{eq:sigkinterms}
\end{equation}
where the gauge covariant derivatives correspond to bi-fundamentals under the gauge groups neighboring the link; for the link field $\Sigma_{ij}$, we have $D_\mu = \partial_\mu - i g_i \hat{A}^i_\mu + i g_j \hat{A}^j_\mu$.

For the scalar higgs interactions, we again impose the $L-R$ symmetry:
\begin{equation}
{\mathcal L}_\text{higgs} = h \left\{ a_h \frac{f_1}{4}   \text{Tr} \left[ \left| D_\mu \Sigma_{L1} \right|^2 \right] +  b_h \frac{f_2}{4}  \text{Tr} \left[ \left| D_\mu \Sigma_{12} \right|^2 \right] +  a_h \frac{f_1}{4} \text{Tr} \left[ \left| D_\mu \Sigma_{2Y} \right|^2 \right] \right\},
\end{equation}
forcing the higgs couplings to the $L-1$ and $2-Y$ kinetic terms to be identical.


\subsection{Couplings in the four site model}
The couplings of the hamiltonian eigenstates, which follow after diagonalization of the quadratic part of the action, can be straightforwardly derived.  Due to the wave-function mixing, however, the normalization condition for the states is modified.  The normalization condition for the eigenvectors in the presence of the wave-function mixing term is instead (as emphasized in~\cite{SekharChivukula:2008gz}) $v_n^T Z v_n = 1$, where we have 
\begin{equation}
Z_0 = \left(\begin{array}{cccc} 1 & 0 & 0 & 0 \\ 0 & 1 & \epsilon & 0 \\ 0 & \epsilon & 1 & 0 \\ 0 & 0 & 0 & 1 \end{array} \right)
\end{equation}
for the neutral gauge bosons and
\begin{equation}
Z_\pm = \left(\begin{array}{ccc} 1 & 0 & 0 \\ 0 & 1 & \epsilon \\ 0 & \epsilon & 1 \end{array} \right)
\end{equation}
for the charged ones.
The eigenvectors thus satisfy the following relation:
\begin{equation}
M^2 \vec{v}_n =  m_n^2 Z \vec{v}_n
\label{eq:eigsol}
\end{equation}
where $M^2$ is the mass matrix of the quadratic lagrangian that follows from the $\Sigma$-field kinetic terms in Eq.~(\ref{eq:sigkinterms}).  To avoid ghost instabilities, we must constrain $\epsilon$ to the interval $-1 < \epsilon < 1$. The components of the eigenvectors, $\vec{v}_n$, are ordered based on the moose structure in Figure~\ref{fig:moosefig}, from left to right.  The couplings of the physical states are then obtained by expressing the original Lagrangian in terms of the eigenvector solutions to Eq.~(\ref{eq:eigsol}).


\subsection{Higgs interactions}
As an example of the interactions of the mass and kinetic eigenstates, we give the Feynman rules for interactions of the scalar higgs with the charged gauge fields in Table~\ref{tab:hint}.
\begin{table}[h]
\centering
\begin{tabular}{|c|c|}
\hline 
$h W^+ W^-$ & $i\frac{2 M_W^2}{v} \left( a_h \frac{s_f^3}{\sqrt{2}} +b_h c_f^3 \right)$ \\
\hline 
$h \rho_V^+ \rho_V^-$ & $i \frac{\sqrt{2} M_\rho^2}{v} a_h s_f$ \\
\hline
$h \rho_A^+ \rho_A^-$ & $i \frac{\sqrt{2} M_A^2}{v} s_f c_f \left( a_h c_f + \sqrt{2} b_h s_f \right)$ \\
   \hline
$h W^+ \rho_A^-$ & $i \frac{2 M_W M_A}{v} s_f c_f \left( a_h \frac{s_f}{\sqrt{2}} - b_h c_f \right)$ \\
   \hline
\end{tabular}
\caption{Feynman rules corresponding to interactions of the singlet field $h$ with charged gauge bosons in the 4-site model shown in Figure~\ref{fig:moosefig}.  We have only kept the lowest order terms in the $\frac{g}{g_\rho}$ expansion;  in fact all interactions are non-vanishing at order $g^2/g_\rho^2$.  The charge reversed Feynman rules are identical.}
\label{tab:hint}
\end{table}
We have performed an expansion in $g/g_\rho$ and $g'/g_\rho$, presuming that the two exotic gauge groups have large (but still perturbative) coupling constants.  We have used the definitions $c_f \equiv f_1/\sqrt{f_1^2+2 f_2^2}$,$s_f \equiv \sqrt{2} f_2/\sqrt{f_1^2+2 f_2^2}$, and $v \equiv f_1 f_2/\sqrt{f_1^2+2f_2^2}$.  When $a_h/f_1\ne b_h/f_2$, the higgs has interactions which change the ``flavor" of gauge field at the vertex.  For models in which  gauge boson self-interactions also allow a change in the flavor of charged gauge boson, a larger class of diagrams than in the SM is allowed.

It is possible that other higher dimensional operators of the form $h V_{\mu\nu}^2$ exist due to strong coupling effects, giving both a direct contribution to higgs decay amplitudes, and also contributing to new loop structures.  We discuss this first possibility later, in Section~\ref{sec:hdocontributions}.  The latter possibility leads to contributions which are suppressed both by loop factors and the cutoff scale.  We neglect such contributions in this work.


\subsection{$\gamma$ and $Z$-boson Interactions with charged spin-1 fields}

The quartic interactions of the photon are constrained by gauge invariance to be simply the electric charge squared, with no ``flavor" changing of the gauge fields at the vertex.  In the $4$-site model under consideration, the quartic Feynman rules are all of the form
\begin{equation}
\lambda_{ij}^{(1)} = \lambda_{ij}^{(2)} = \lambda_{ij}^{(3)} = \left\{ \begin{array}{cl} e^2 & i = j \\ 0 & i \ne j \end{array} \right.
\end{equation}
where the electric charge is given in terms of the fundamental model parameters as
\begin{equation}
e^2 = e_0^2 \left( 1 - \frac{2 e_0^2 \left(1+\epsilon \right) }{g_\rho^2} + {\mathcal O} (e_0^4/g_\rho^4)\right),
\label{eq:chargeexp}
\end{equation}
with $e_0^2 \equiv g^2 g'^2/(g^2+g'^2)$.

For the cubic interactions, the presence of the wave function mixing term induces off-diagonal couplings of the photon to charged spin-1 fields.  The interactions, to order $1/g_\rho^2$, are given in Table~\ref{tab:phoint}.

\begin{table}[h]
\centering
\begin{tabular}{|c|c|c|}
\hline
\multirow{3}{*}{$\gamma W^+ W^-$}
&$g_0$ & $e\left( 1 + \epsilon c_f^4 \left( \frac{g}{g_\rho} \right)^2 \right)$ \\ \cline{2-3}
&$g_+$ & $e$ \\ \cline{2-3}
&$g_-$ & $e$ \\ \hline
\multirow{1}{*}{$\gamma W^+ \rho_A^-$}
&$g_0$ & $e \epsilon  c_f^2 \sqrt{\frac{2}{1-\epsilon}}\left( \frac{g}{g_\rho} \right)$ \\ \hline
\multirow{3}{*}{$\gamma \rho_V^+ \rho_V^-$}
&$g_0$ & $e$ \\ \cline{2-3}
&$g_+$ & $e$ \\ \cline{2-3}
&$g_-$ & $e$ \\ \hline
\multirow{1}{*}{$\gamma \rho_V^+ \rho_A^-$}
&$g_0$ & $e \epsilon  c_f^2 \left(1+ \epsilon \right) \sqrt{\frac{1+\epsilon}{1-\epsilon}} \frac{1}{2 \left( \epsilon c_f^2 + \frac{1}{2}\left(1+ \epsilon\right)s_f^2 \right)} \left( \frac{g}{g_\rho} \right)^2$      \\ \hline
\multirow{3}{*}{$\gamma \rho_A^+ \rho_A^-$}
&$g_0$ & $e \left( \frac{1+\epsilon}{1-\epsilon} - \epsilon c_f^4\left( \frac{g}{g_\rho} \right)^2 \right)$ \\ \cline{2-3}
&$g_+$ & $e$ \\ \cline{2-3}
&$g_-$ & $e$ \\ \hline
\end{tabular}
\caption{These are the non-vanishing (at order $g/g_\rho^2$) cubic interactions of the photon with the charged gauge bosons associated with the moose diagram in Figure~\ref{fig:moosefig}.  The expression for $e$ in terms of the fundamental parameters (to order $g^2/g_\rho^2$) is given in Eq.~\ref{eq:chargeexp}.}
\label{tab:phoint}
\end{table}

The corresponding interactions of the $Z$-boson with charged spin-1 fields are algebraically much more complicated.  We have made the full set of couplings, valid to order $g^2/g_\rho^2$, available as Mathematica code~\cite{mathematicafiles}.


\subsection{Loop level contributions to $h \rightarrow Z \gamma$ and $h \rightarrow \gamma \gamma$}
\label{sec:loopresults}
The following results are analytic expressions that are valid to lowest order in the $g/g_\rho$ expansion, and correspond to the low energy theorem limit where $m_h,m_Z \ll 2 m_{V^\pm}$.  The amplitudes are proportional to the usual transverse tensor structure 
\begin{align}
e^2 \left( g^{\alpha_1 \alpha_2} p_{\gamma^1} \cdot p_{\gamma^2} - p_{\gamma^1}^{\alpha_1} p_{\gamma^2}^{\alpha_2} \right) \mathcal{M}_{\gamma\gamma} \nonumber \\
e g \cos \theta_w \left( g^{\alpha_1 \alpha_2} p_{Z} \cdot p_{\gamma} - p_{Z}^{\alpha_1} p_{\gamma}^{\alpha_2} \right) \mathcal{M}_{Z\gamma} 
\end{align}
For the 4-site model, performing the summations of Eq.~\ref{eq:ampsum}, we find:
\begin{align}
&\mathcal{M}_{\gamma\gamma} = \frac{\epsilon^2 f_1\log \frac{\Lambda^2}{M_A^2} }{4 \pi^2 f_2^3 (1-\epsilon)^2} \left( a_h f_2 - b_h f_1 \right) + \frac{7}{8 \pi^2 f_1 f_2} \left( 2 a_h f_2 + b_h f_1 \right)\nonumber\\
& + \frac{\epsilon}{8 \pi^2 f_2 \left(1-\epsilon\right)^2} \left[ a_h \sqrt{2} \frac{c_f}{s_f} \left( c_f^2 \epsilon + 3 s_f^2 (2-\epsilon) \right)+b_h \left( 12 s_f^2 (1-\epsilon)-10 c_f^2 \epsilon - 6 \frac{c_f^4}{s_f^2} \epsilon \right) \right]  \\
~ & \nonumber \\
&\mathcal{M}_{Z\gamma} = \frac{\epsilon \log \frac{\Lambda^2}{M_A^2}}{2 \sqrt{2} \pi^2 (1-\epsilon)^2} \frac{c_f}{s_f^3} \frac{\left( a_h f_2 - b_h f_1 \right)}{f_1^2+2 f_2^2} \left[ \epsilon \left( 1- \tan^2 \theta_w \right) - \frac{1}{2}  s_f^2 \left( 1- \epsilon \right) \left(1+ \tan^2 \theta_w\right) \right] \nonumber \\
& + \frac{7 }{16 \pi^2 f_1 f_2} \left[ \left( 2 a_h f_2 + b_h f_1 \right) \left( 1 - \tan^2 \theta_w \right) +   \left(  a_h f_2 s_f^2 + b_h f_1 c_f^2 \right) \left(1+ \tan^2 \theta_w \right) \right] \nonumber \\
& + \frac{\epsilon}{16 \pi^2 (1-\epsilon)^2} \frac{f_1}{f_2^2} \left\{ a_h \left[ \left( 3 s_f^2 (2-\epsilon) + \epsilon c_f^2\right)  \left( 1 - \tan^2 \theta_w \right)- \frac{3}{2} s_f^2 (1- \epsilon) \left( 1+ \tan^2 \theta_w \right) \right]  \right. \nonumber \\
& \left. + b_h \left[\left( 6 \sqrt{2} \frac{s_f^3}{c_f} (1-\epsilon) - 2 \sqrt{2} c_f s_f \epsilon - 3 \sqrt{2} \frac{c_f}{s_f}\epsilon \right) \left(1- \tan^2 \theta_w \right)+ \frac{3}{\sqrt{2}} s_f c_f (1-\epsilon) \left( 1+ \tan^2 \theta_w \right) \right] \right\}.
\end{align}

In the $\epsilon \rightarrow 0$ limit, when the non-renormalizable operator incorporating wave-function mixing is turned off, the results are finite and given by
\begin{align}
&\mathcal{M}_{\gamma\gamma} = \frac{7}{8 \pi^2 f_1 f_2} \left( 2 a_h f_2 + b_h f_1 \right) \\
~ & \nonumber \\
&\mathcal{M}_{Z\gamma} =  \frac{7 }{16 \pi^2 f_1 f_2} \left[ \left( 2 a_h f_2 + b_h f_1 \right) \left( 1 - \tan^2 \theta_w \right) +   \left(  a_h f_2 s_f^2 + b_h f_1 c_f^2 \right) \left(1+ \tan^2 \theta_w \right) \right].
\end{align}
For our numerical analysis, we use these formulae to compare against the standard model expectations for these amplitudes.


\subsection{$h \rightarrow Z \gamma$ and $h \rightarrow \gamma \gamma$ from higher dimensional operators}
\label{sec:hdocontributions}
There are tree-level contributions to the $h Z \gamma$ and $h \gamma \gamma$ couplings inherited from strong-coupling effects~\cite{Azatov:2013ura}  that couple the scalar $h$ directly to the field strengths of the two middle $SU(2)$ groups in the moose.  These terms serve as counter-terms for divergences that appear in loop amplitudes such as those given in the previous sub-section.  We have not considered tree level couplings to the ``fundamental" $W$ and $Z$ bosons in the effective field theory (the gauge groups on either end of the moose) but since there is mixing after symmetry breaking takes place, there is an effective tree level $h Z \gamma$ coupling.  The tree level L-R symmetric lagrangian before spontaneous breaking is assumed to take the form
\begin{equation}
\frac{c}{4 \Lambda} h \left[ \left( \rho_1^{\mu\nu~a} \right)^2+\left( \rho_2^{\mu\nu~a} \right)^2 \right] + \frac{c_\epsilon}{2\Lambda} h \text{Tr} \left[ \rho_{1~\mu\nu} \Sigma_{12} \rho_2^{\mu\nu}\Sigma_{12}^\dagger \right]
\label{eq:hWF}                                                                                                                                                                                       
\end{equation}
where $c$ is an unknown coefficient parametrizing the effects of UV strongly-coupled dynamics. Like the WF mixing term, $c_\epsilon$ is an addtional coefficient parametrizing the ``mixing'' between the two heavy vectors.  After the theory is expressed in the mass basis, the resulting lagrangian term is
\begin{equation}
\frac{\left( c +c_\epsilon\right)}{2\Lambda}\frac{ e g \cos \theta_w}{g_\rho^2} \left(1-\tan^2\theta_w \right)  h Z_{\mu\nu} A^{\mu \nu} +\frac{\left(c +c_\epsilon\right)}{2\Lambda} \frac{e^2}{g_\rho^2} h A_{\mu\nu} A^{\mu \nu}.
\end{equation}

Note that the additional contributions to the amplitudes both scale in the same way, and with the same sign as functions of the coefficients $c$ and $c_\epsilon$.  Generically, strong coupling effects are expected to produce values of the $c$ parameters that are of order $g_\rho^2$, such that these terms serve as counter-terms to absorb the divergences in the amplitudes for the higgs decay rates.


\section{Results}
\label{sec:results}
Current LHC constraints on the higgs couplings favor a SM-like coupling of the higgs to $Z$ bosons.  The $hW^+W^-$ coupling is given in Table~\ref{tab:hint}, and the $hZZ$ coupling is of a similar form.  To leading order in the $g/g_\rho$ expansion, for massive vectors $V$, we have
\begin{equation}
\frac{g_{h VV}}{g_{hVV}^\text{SM}} = a_h \frac{s_f^3}{\sqrt{2}}  + b_h c_f^3 + \mathcal{O}(g^2/g_\rho^2).
\end{equation}
In plotting our results, we constrain $g_{h VV}/g_{hVV}^\text{SM} = 1$, enforcing a relationship between $a_h$ and $b_h$.  We choose to eliminate $b_h$ with this relation, and vary $a_h$.  We have also restricted the $W$-boson mass to the SM value, fixing one combination of $f_1$ and $f_2$.  To leading order in $g/g_\rho$, we set
\begin{equation}
v^2 = \frac{f_1^2 f_2^2}{f_1^2+2 f_2^2},
\end{equation}
with $v \equiv 246$ GeV.

\begin{figure}[t]
\centering
\includegraphics[width=.6\hsize]{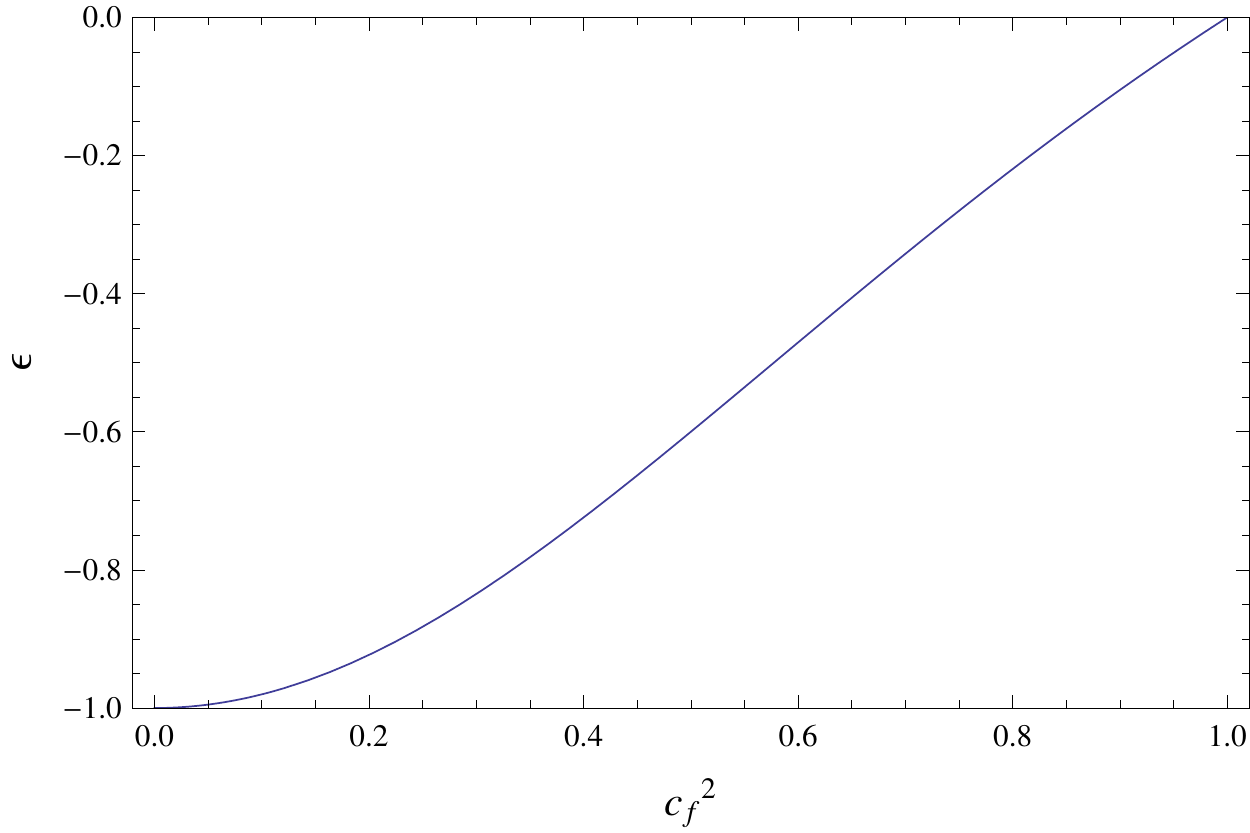}
\caption{Values of $\epsilon$ for which the $S$ parameter vanishes as a function of the angle $c_f^2 \equiv f_1^2/(f_1^2+2 f_2^2)$.  The high and low ranges of $c_f$ correspond to large hierarchies between the vev's $f_1$ and $f_2$.  The large $c_f$ limit, in which $f_1 \rightarrow \infty$, is the decoupling limit for the vector and axial vector.}
\label{fig:Siszero}
\end{figure}

One of the motivations for considering an effective field theory which contains an axial vector is to study the interplay between electroweak precision and the higgs decay rates.  In strongly coupled models of electroweak symmetry breaking or in their holographic counterparts, ameliorating the tree-level contributions to $S$ is a particular challenge~\cite{Agashe:2007mc,Cacciapaglia:2004rb,Cacciapaglia:2006gp}.  In adjusting the parameter $\epsilon$, the relationship between the axial-vector and vector resonances change:
\begin{equation}
\frac{M_\rho^2}{M_A^2} = c_f^2 \frac{1-\epsilon}{1+\epsilon} + \mathcal{O}(g^2/g_\rho^2),
\end{equation}
and the tree-level contributions to the $S$ parameter vary accordingly~\cite{SekharChivukula:2008gz}:
\begin{equation}
\Delta S\approx \frac{2 \sin^2 \theta_w}{\alpha} \frac{g^2}{g_\rho^2} \left(1+\epsilon\right) \left( 1- c_f^4 \frac{1-\epsilon}{1+\epsilon}  \right) \approx \frac{4 \sin^2 \theta_w}{\alpha s_f^2} \frac{M_W^2}{M_\rho^2} \left( 1- c_f^2 \frac{M_\rho^2}{M_A^2}  \right).
\end{equation}
In Figure~\ref{fig:Siszero}, we show the value of $\epsilon$ that is required for the tree level value of $S$ to be zero.  We note that negative $\mathcal{O}(1)$ values must be taken to completely set $S$ to zero for a large range of $c_f$.  It is only in the $c_f \rightarrow 1$ limit that only small values of epsilon are necessary.  However, that limit corresponds precisely to the decoupling limit $f_1\rightarrow \infty$, in which both the vector and axial vector masses become large.

For $\mathcal{O}(1)$ negative values of $\epsilon$, the normal hierarchy between the vector and axial vector resonances is inverted, and the S-parameter can be reduced to zero.  Note, however, that such large values of $\epsilon$ exceed expectations from application of naive dimensional analysis~\cite{Georgi:1992dw}, and there are arguments against such an inverted spectrum following from studies of holographic technicolor models~\cite{Agashe:2007mc}.   For the purposes of this work, however, we are motivated more by phenomenological exploration.  For example, one question of merit is whether there exists a correlation between values of $S$ and loop corrections to the $h\rightarrow \gamma \gamma (Z\gamma)$ rates that may persist generically in more realistic models of electroweak symmetry breaking.  In this spirit, we display results for ranges of $\epsilon$ following only the requirements that the theory remain perturbative and that the spectrum be tachyon-/ghost-free.

\begin{figure}[h!]
\centering
\includegraphics[width=.32\hsize]{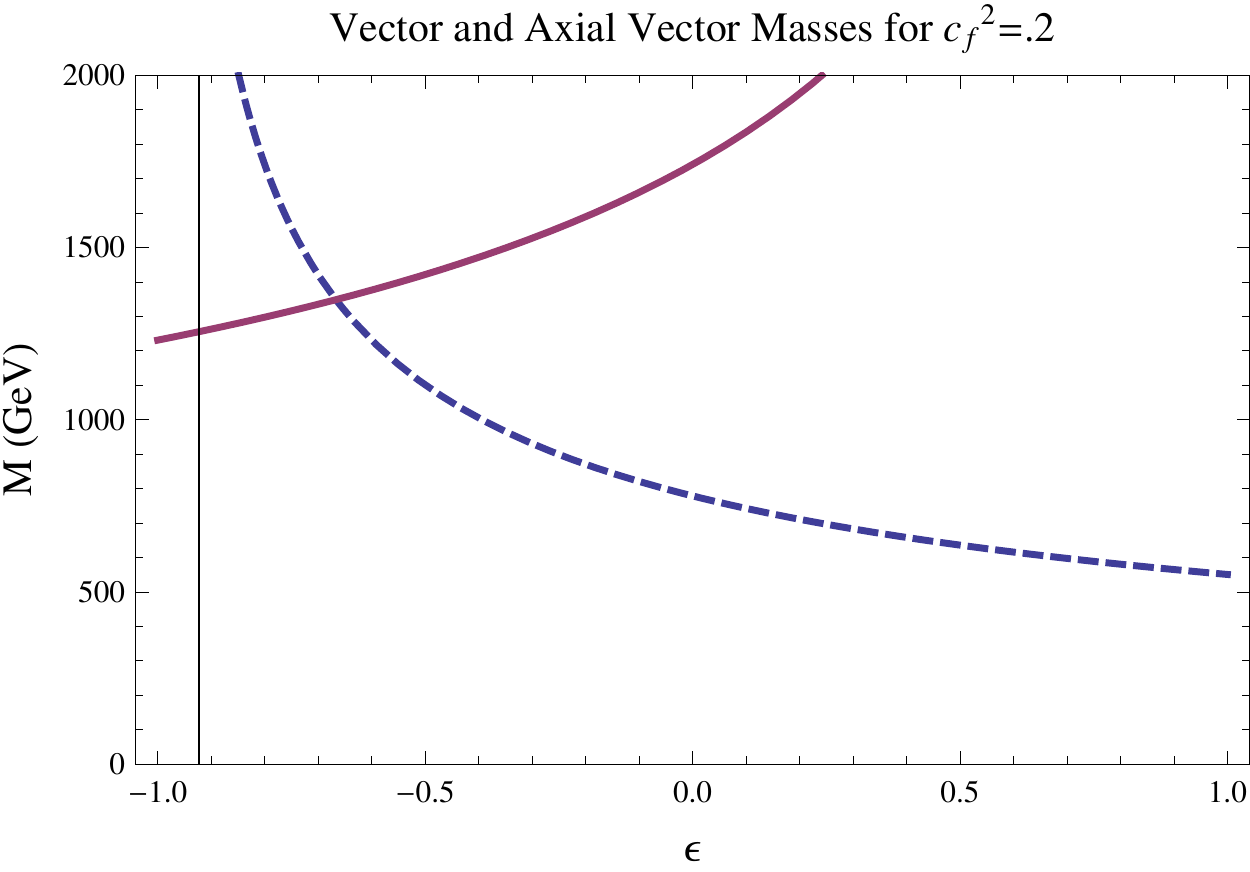}
\includegraphics[width=.32\hsize]{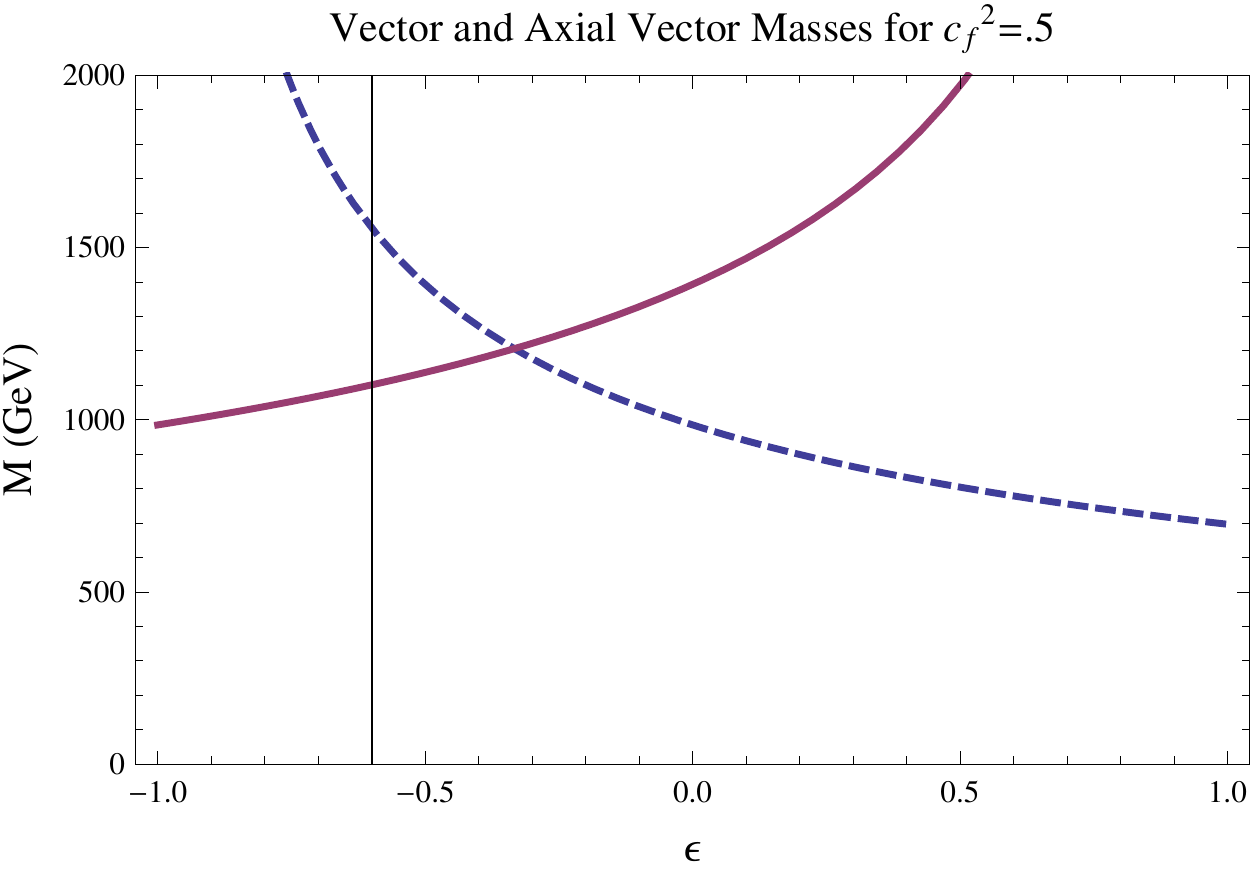}
\includegraphics[width=.32\hsize]{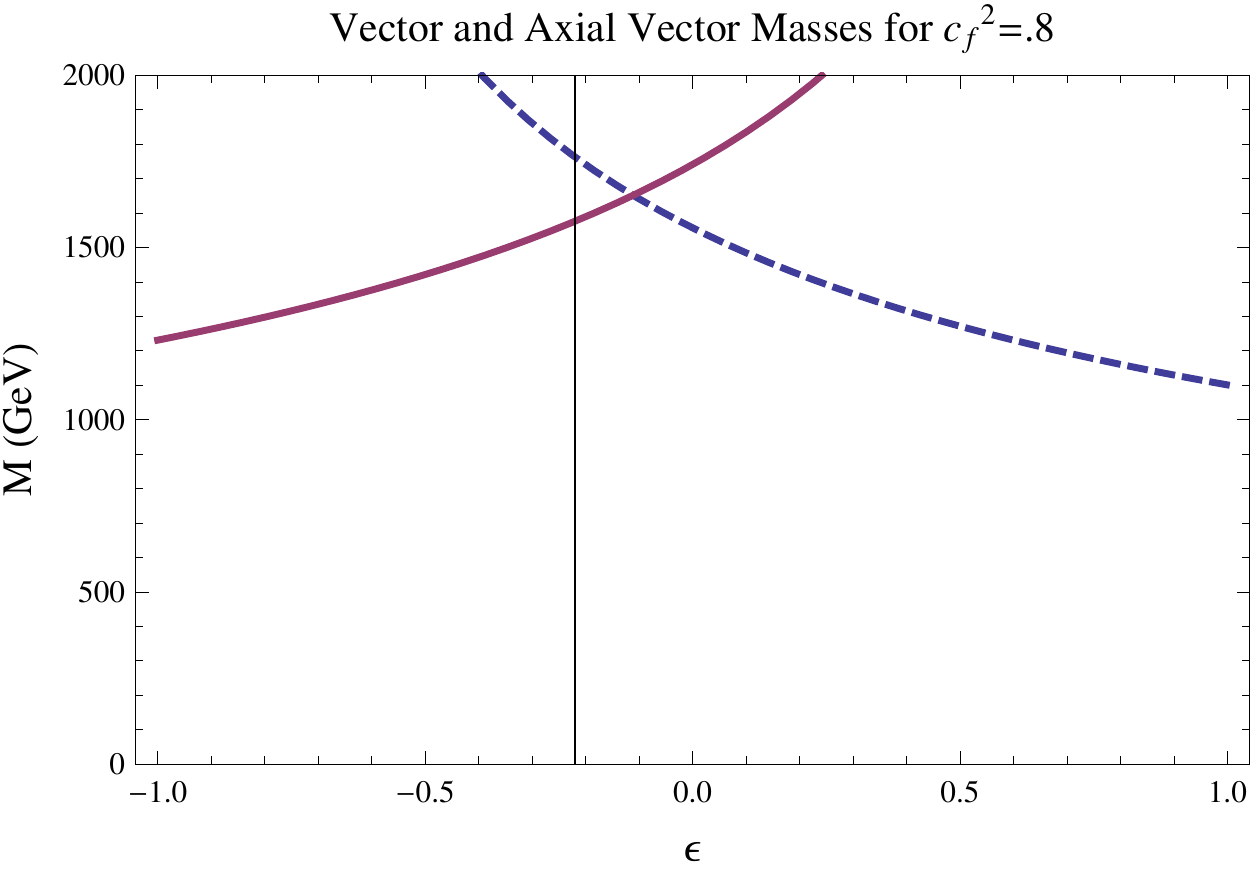}
\caption{The values of $M_\rho$ (dashed) and $M_A$ (solid) as a function of $\epsilon$ for $c_f^2 = .2$, $.5$, and $.8$, respectively.  The value of $g_\rho$ has been fixed at $g_\rho = 4$ in this figure, however the masses scale linearly with $g_\rho$, so long as it is large compared with electroweak gauge couplings.  The black vertical lines correspond to the values of $\epsilon$ for which the tree-level contribution to $S$ vanishes.}
\label{fig:VandAVmasses}
\end{figure}

In Figure~\ref{fig:VandAVmasses}, we display the vector and axial vector masses (for the choice $g_\rho=4$) as a function of $\epsilon$ for various choices of $c_f$.  The black vertical lines display the value of $\epsilon$ for which the tree-level S-parameter vanishes.  Note that the inverted spectrum is required for the tree-level contribution to $S$ to vanish.  The leading order (in the $g/g_\rho$ expansion) expressions for the vector and axial vector masses are given by
\begin{align}
&M_\rho^2 = \frac{g_\rho^2 f_1^2}{4 (1+ \epsilon)} \nonumber \\
&M_A^2 = \frac{g_\rho^2 (f_1^2+2 f_2^2)}{4(1-\epsilon)}.
\end{align}
 
\begin{figure}[h!]
\centering
\includegraphics[width=.49\hsize]{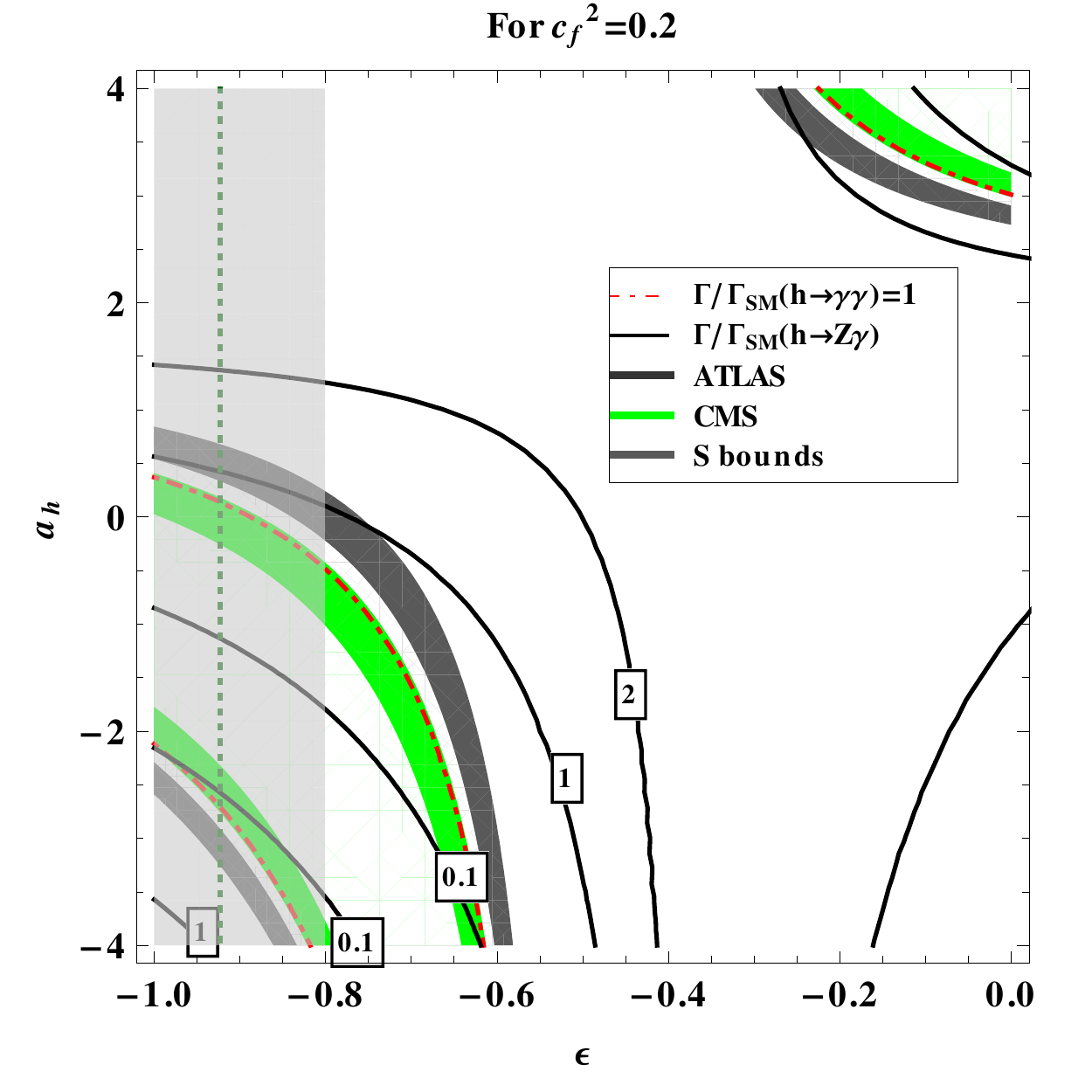}
\includegraphics[width=.49\hsize]{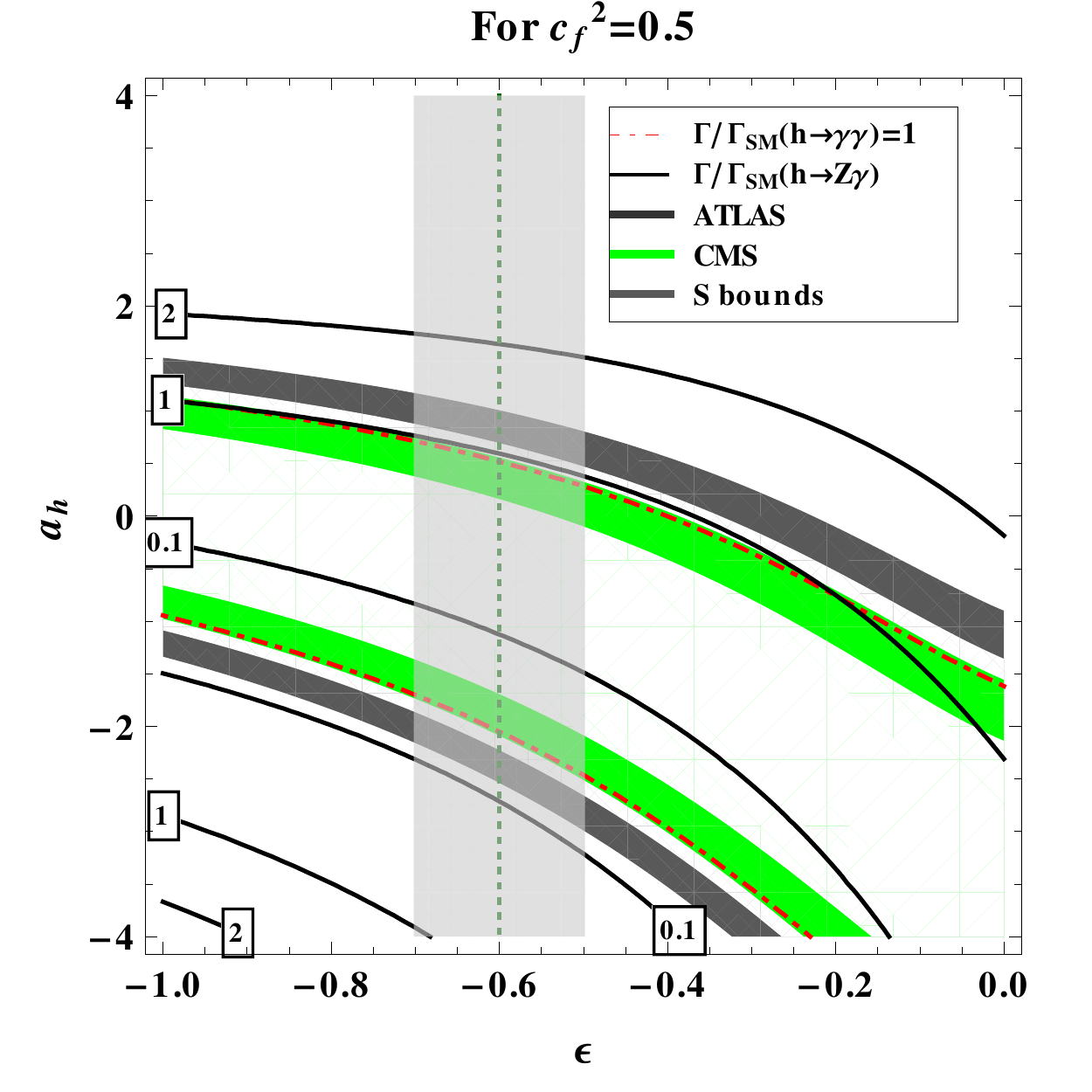}
\includegraphics[width=.49\hsize]{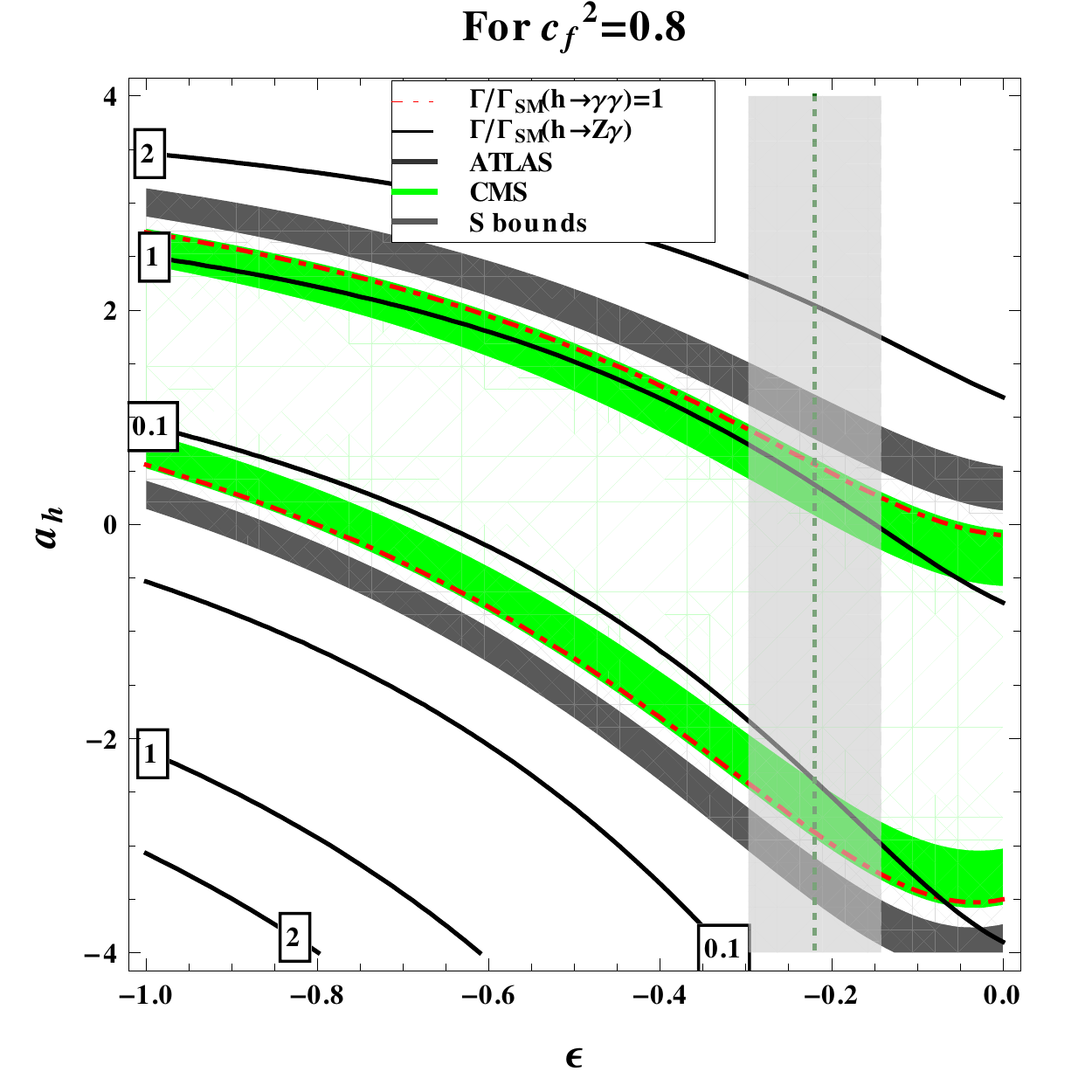}
\includegraphics[width=.49\hsize]{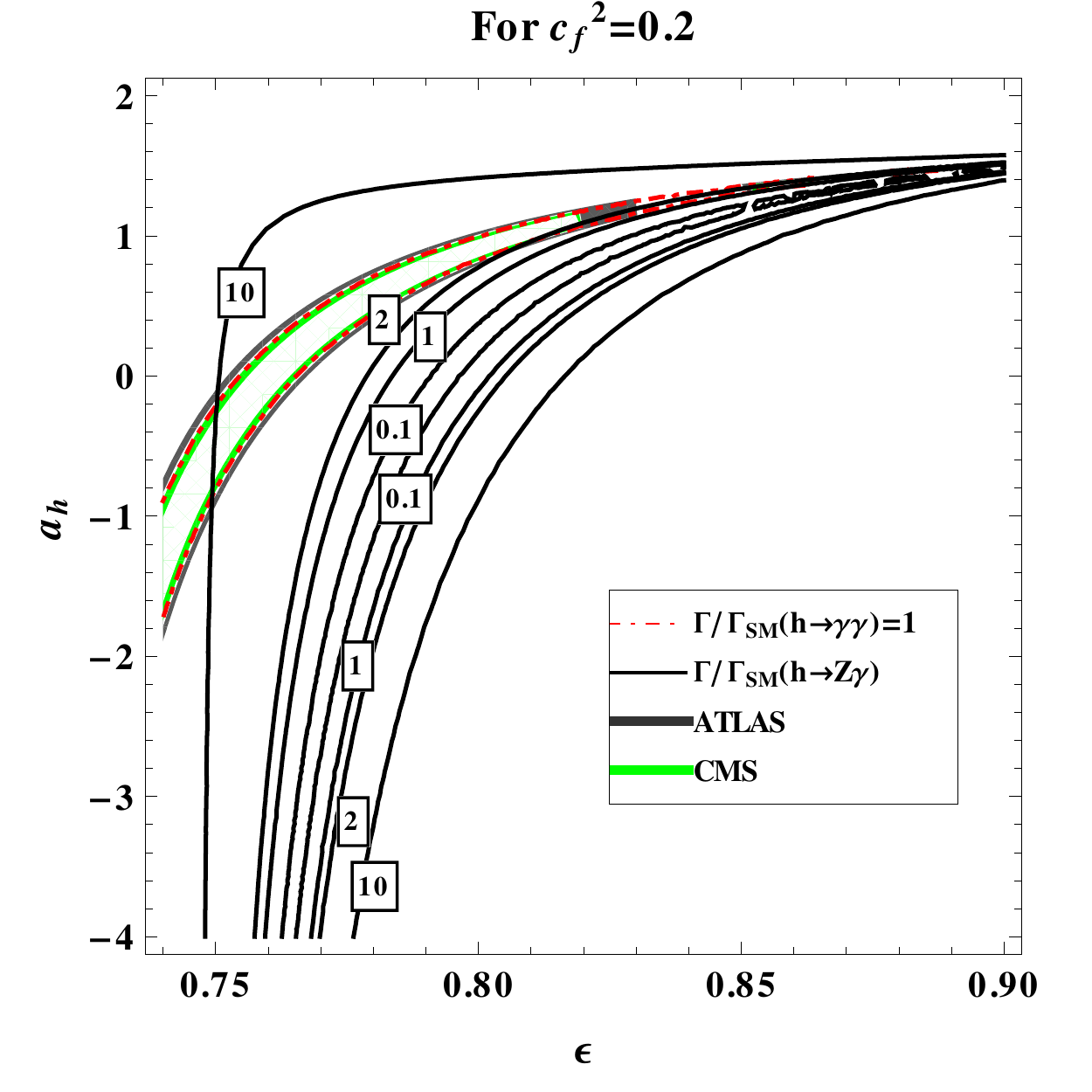}
\caption{This figure displays the ratio of the higgs partial widths to the $Z\gamma$ and $\gamma\gamma$ final states in relation to the expectation in the SM.  The figures represent the scenario where direct contributions from higher dimensional operators are neglected.  Loop diagrams from the vector and axial vector states are taken into account.   The three plots are for $c_f^2=0.2,0.5, \text{ and } 0.8$.  The light grey shaded region corresponds to the value of $\epsilon$ for which the $S$-parameter obeys current experimental constraints~\cite{Baak:2012kk}.  The dark grey and green bands correspond respectively to the $1\sigma$ bands for the ATLAS~\cite{Aad:2013wqa} and CMS~\cite{CMSgamgam} experimental results for $h \rightarrow \gamma\gamma$.}
\label{fig:decaywidths}
\end{figure}
\begin{figure}[h]
\centering
\includegraphics[width=.49\hsize]{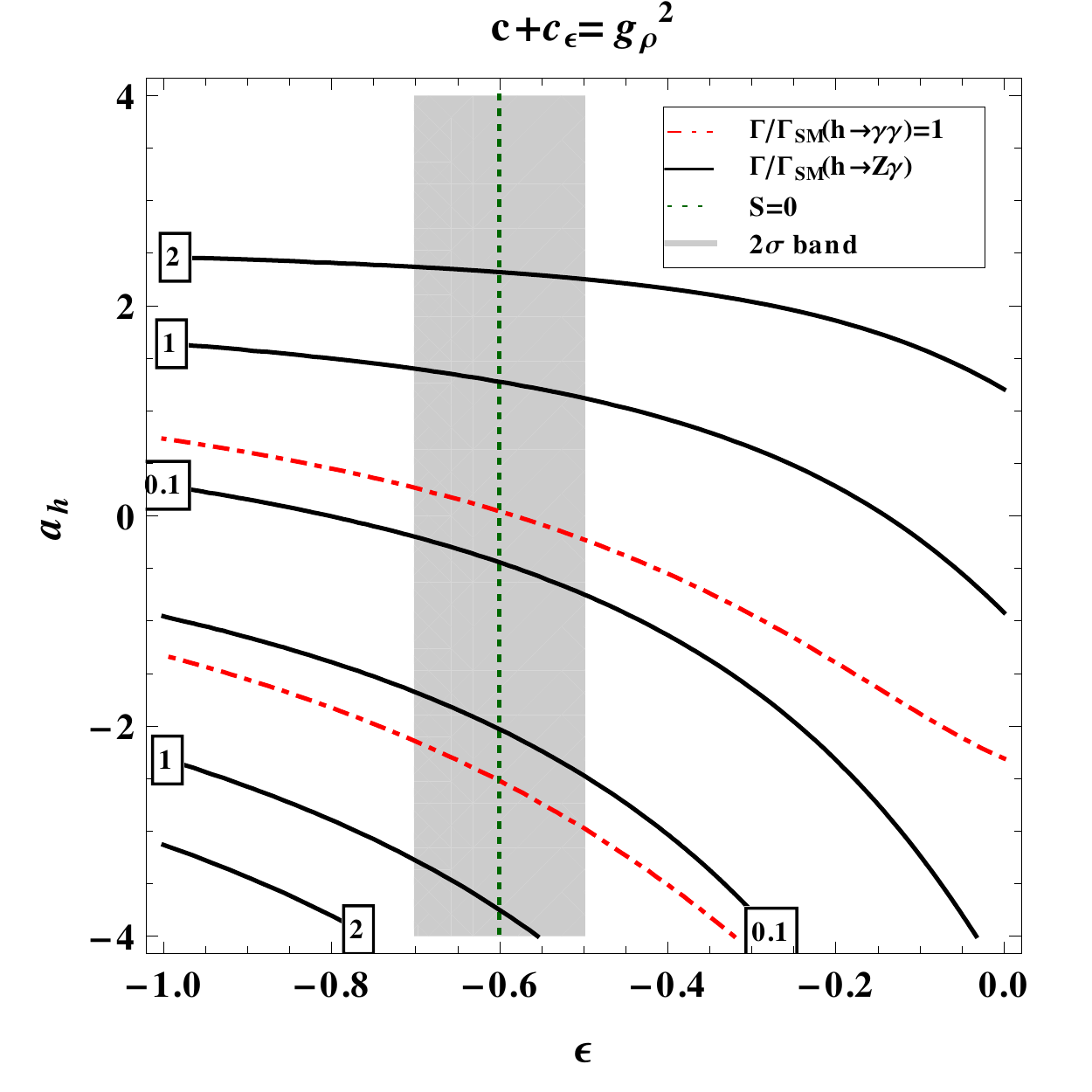}
\includegraphics[width=.49\hsize]{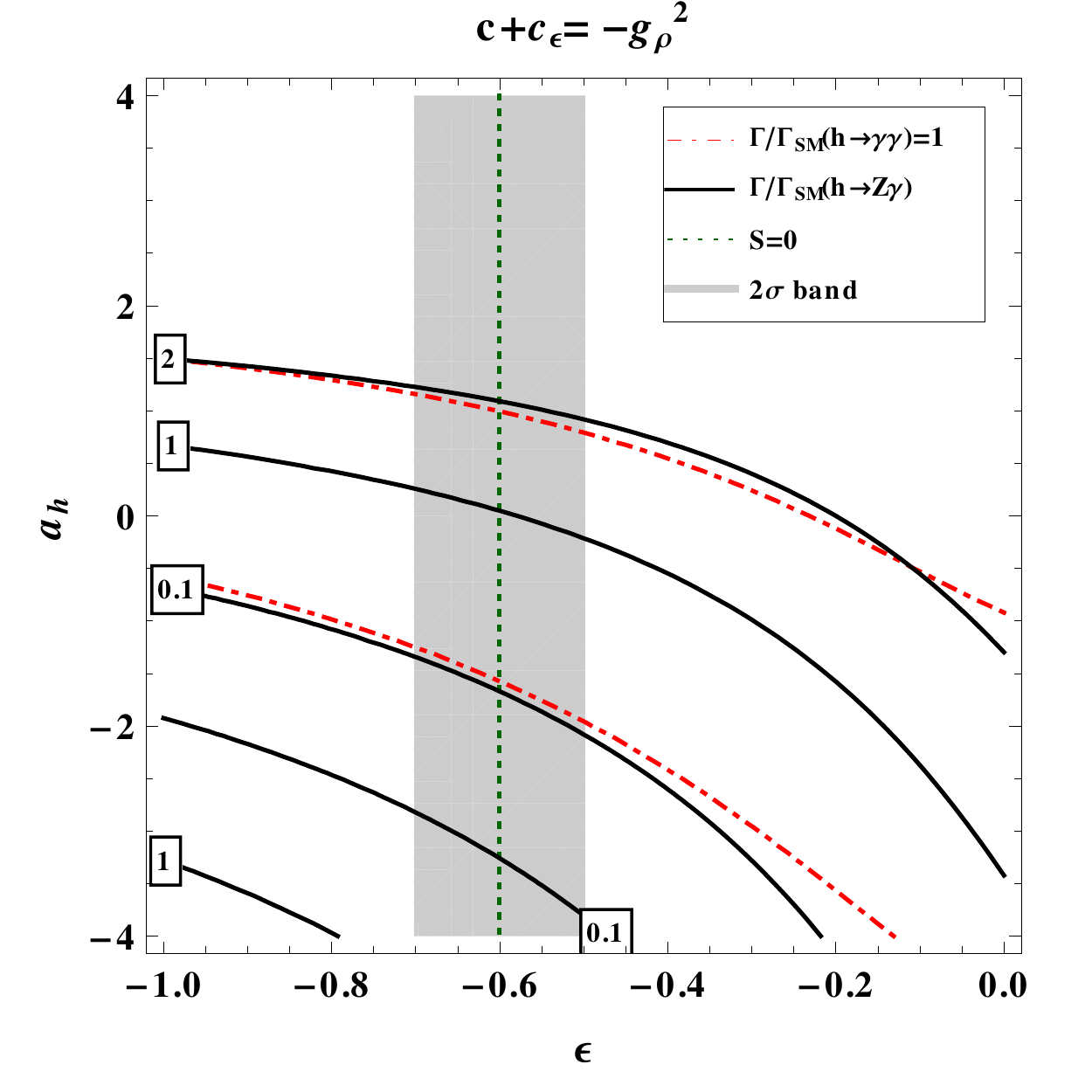}
\caption{This figure displays the ratio of the higgs partial widths to the $Z\gamma$ and $\gamma\gamma$ final states in relation to the expectation in the standard model when dimension 5 operators coupling the higgs field directly to exotic field strengths are added, interfering with the loop level contributions of states in the low energy effective theory.  In the two plots, we have taken $c+c_\epsilon = g_\rho^2$ (left) and $c+c_\epsilon = -g_\rho^2$ (right).  For these plots, we have taken $c_f^2 = 0.5$.  We have fixed the cutoff scale $\Lambda$ at twice the mass of the axial-vector resonance, which varies as a function of $\epsilon$ and $c_f$ as shown in Figure~\ref{fig:VandAVmasses}.}
\label{fig:hdofig}
\end{figure}

We have added the amplitudes calculated in Section~\ref{sec:loopresults} to the SM top quark contributions for both $h\rightarrow \gamma \gamma$ and $h\rightarrow Z \gamma$, and calculated the partial decay widths to these final states in the 4-site model.  Comparing with the SM rates\footnote{We utilize the higgs low energy theorem limits for both the SM W contribution and the new physics contribution to make this comparison.}, we display the ratio $\Gamma (h \rightarrow XX)/\Gamma_\text{SM} (h \rightarrow XX)$ as functions of $\epsilon$ and $a_h$ for 3 representative values of the angle $c_f$.  While significant enhancements or suppressions are possible in the theory, we find that when the $h \rightarrow \gamma \gamma$ rate is SM-like (as suggested by current LHC data), the contributions to $h \rightarrow Z\gamma$ are either close to SM-like as well, or experience a large suppression (where the branching ratio is approximately 1/10'th that of the SM).  In Figure~\ref{fig:decaywidths}, we display the decay rates in the branching fractions for these final states relative to SM expectations.  In this plot we have taken the contributions from the higher dimensional operators discussed in Section~\ref{sec:hdocontributions} to be vanishing (i.e. $c=c_\epsilon = 0$).  Adding these operators with non-trivial coefficients changes the contour bands, as Shown in Figure~\ref{fig:hdofig}.  In both Figure~\ref{fig:decaywidths} and Figure~\ref{fig:hdofig}, we have kept the ratio of the cutoff scale and the axial vector masses fixed at $\Lambda/M_A = 2$.  Since $M_A$ varies with $\epsilon$ and $c_f$, the cutoff changes in these plots as well.  While the shape of the contours does not change significantly with the addition of these operators, it is important to note that the relative size of the $h\rightarrow \gamma \gamma$ rates vs the $h\rightarrow Z \gamma$ rates differ significantly.  For example, with the higher dimensional operator coefficients set to zero, there is mostly only a suppression of the $h\rightarrow Z\gamma$ rates when the $\gamma \gamma$ rate is SM-like.  In contrast, when the higher dimensional operators are added with coefficients consistent with naive dimensional analysis, the $Z\gamma$ rate can either be signficantly suppressed relative to SM predictions (see left panel in Figure~\ref{fig:hdofig}), or potentially enhanced (see right panel in Figure~\ref{fig:hdofig}) depending on the sign of their coefficients.


\section{Conclusions}
\label{sec:conclusions}
We have considered the effects of electroweak/TeV scale spin-1 resonances on the phenomenology of a higgs-like scalar resonance.  In particular, we have calculated the effects of such fields on the di-boson decays:  $h \rightarrow \gamma \gamma$ and $h \rightarrow Z\gamma$.  A very general framework for calculations of spin-1 contributions has been constructed, with application to arbitrary gauge extensions of the SM made possible via Mathematica files that have been made available online~\cite{mathematicafiles}.  In these files, the quantum effects of vector-resonances and SM gauge fields have been presented as functions of generic couplings that may arise in extra dimensional models, little higgs models, strongly coupled theories, or various other SM extensions with exotic spin-1 resonances that couple to the electroweak sector.

The results of this calculation have been applied to a benchmark phenomenological model for dynamical electroweak symmetry breaking that contains a composite scalar resonance in the spectrum.  In particular, the effects of a class of models with vector and axial-vector triplets on scalar phenomenology have been computed and found to generate potentially large contributions to the $\gamma\gamma$ and $\gamma Z$ branching fractions of the $125$ GeV resonance.  Contributions to the higgs decay rates are especially interesting in these scenarios, as the divergence structure of the decay amplitudes is dependent on the value of the parameter that determines the size of tree-level contributions to the S-parameter.

Future runs of the LHC, including both energy and luminosity upgrades, are likely to strongly constrain the viability of many gauge extensions of the SM via probes of the higgs, particularly once we measure its decay rate to the $Z\gamma$ final state.  The correlations of this channel with electroweak precision constraints and the $h\rightarrow \gamma \gamma$ rate are particularly interesting in light of the current state of the allowed landscape of well-motivated gauge extensions of the SM.  We have provided here a set of tools which we hope will be a valuable resource as we test such theories against LHC data.


\section*{Acknowledgements}

We thank Brando Bellazzini and Csaba Cs\'aki for helpful discussions while this work was in progress.  D.B., J.H., and B.J.~thank Cornell University for hospitality throughout the course of this work.  D.B.~thanks Syracuse University for hospitality following his thesis defense.  J.H. thanks the Kavli Institute for Theoretical Physics at U.C. Santa Barbara~for hospitality while this work was being completed.   J.H. was supported by the DOE under grant DE-FG02-85ER40237.



\end{document}